\documentclass[aps,prc,twocolumn,superscriptaddress,showpacs]{revtex4-1}
\usepackage[dvips]{graphicx}
\usepackage{bm,color}
\usepackage{dcolumn}
\usepackage{amsmath}

\def\lamb#1#2{$^{#1}_{\Lambda}${#2}} % A < 10
\def\lam#1#2{$^{#1}_{~\Lambda}${#2}} % A .ge. 10
\def\Kpig{($K^-,\pi^- \gamma$) }
\def\piKg{($\pi^+,K^+\gamma$) }
\def\etal{\textit{et al.}}

\begin{document}

\title {Spectroscopy of $^{\bm {9}}_{\bm{\Lambda}}$Li by
electroproduction}

\author{G.M.~Urciuoli}
\affiliation{Istituto Nazionale di Fisica Nucleare, Sezione di Roma,
Piazzale A. Moro 2, I-00185 Rome, Italy}

\author{F.~Cusanno}\thanks{Deceased}
\affiliation{Istituto Nazionale di Fisica Nucleare, Sezione di Roma,
Piazzale A. Moro 2, I-00185 Rome, Italy}

\author{S.~Marrone}
\affiliation{Istituto Nazionale di Fisica Nucleare, Sezione di Bari and
University of Bari, I-70126 Bari, Italy}

\author{A.~Acha}
\affiliation{Florida International University, Miami, Florida 33199, USA}

\author{P.~Ambrozewicz}
\affiliation{Florida International University, Miami, Florida 33199, USA}

\author{K.A.~Aniol}
\affiliation{California State University, Los Angeles, Los Angeles
California 90032, USA}

\author{P.~Baturin}
\affiliation{Florida International University, Miami, Florida 33199, USA}

\author{P.Y.~Bertin}
\affiliation{Universit\'{e} Blaise Pascal/IN2P3, F-63177 Aubi\`{e}re, France}

\author{H.~Benaoum}
\affiliation{Department of Applied Physics, University of Sharjah, UAE}

\author{K.I.~Blomqvist}
\affiliation{Universit\"at Mainz, Mainz, Germany}

\author{W.U.~Boeglin}
\affiliation{Florida International University, Miami, Florida 33199, USA}

\author{H.~Breuer}
\affiliation{University of Maryland, College Park, Maryland 20742, USA}

\author{P.~Brindza}
\affiliation{Thomas Jefferson National Accelerator Facility, Newport News,
Virginia 23606, USA}

\author{P.~Byd\v{z}ovsk\'y}
\affiliation{Nuclear Physics Institute, \v{R}e\v{z} near Prague, Czech
Republic}

\author{A.~Camsonne}
\affiliation{Universit\'{e} Blaise Pascal/IN2P3, F-63177 Aubi\`{e}re, France}

\author{C.C.~Chang}
\affiliation{University of Maryland, College Park, Maryland 20742, USA}

\author{J.-P.~Chen}
\affiliation{Thomas Jefferson National Accelerator Facility, Newport News,
Virginia 23606, USA}

\author{Seonho~Choi}
\affiliation{Temple University, Philadelphia, Pennsylvania 19122, USA}

\author{E.A.~Chudakov}
\affiliation{Thomas Jefferson National Accelerator Facility, Newport News,
Virginia 23606, USA}

\author{E.~Cisbani}

\author{S.~Colilli}
\affiliation{Istituto Nazionale di Fisica Nucleare, Sezione di Roma, gruppo
collegato Sanit\`a, and Istituto Superiore di Sanit\`a, I-00161 Rome, Italy}

\author{L.~Coman}
\affiliation{Florida International University, Miami, Florida 33199, USA}

\author{B.J.~Craver}
\affiliation{University of Virginia, Charlottesville, Virginia 22904, USA}

\author{G.~De~Cataldo}
\affiliation{Istituto Nazionale di Fisica Nucleare, Sezione di Bari and
University of Bari, I-70126 Bari, Italy}

\author{C.W.~de~Jager}
\affiliation{Thomas Jefferson National Accelerator Facility, Newport News,
Virginia 23606, USA}, \affiliation{University of Virginia, Charlottesville, Virginia 22904, USA}

\author{R.~De~Leo}
\affiliation{Istituto Nazionale di Fisica Nucleare, Sezione di Bari and
University of Bari, I-70126 Bari, Italy}

\author{A.P.~Deur}
\affiliation{University of Virginia, Charlottesville, Virginia 22904, USA}

\author{C.~Ferdi}
\affiliation{Universit\'{e} Blaise Pascal/IN2P3, F-63177 Aubi\`{e}re, France}

\author{R.J.~Feuerbach}
\affiliation{Thomas Jefferson National Accelerator Facility, Newport News,
Virginia 23606, USA}

\author{E.~Folts}
\affiliation{Thomas Jefferson National Accelerator Facility, Newport News,
Virginia 23606, USA}

\author{R.~Fratoni}

\author{S.~Frullani}

\author{F.~Garibaldi}
\affiliation{Istituto Nazionale di Fisica Nucleare, Sezione di Roma, gruppo
collegato Sanit\`a, and Istituto Superiore di Sanit\`a, I-00161 Rome, Italy}

\author{O.~Gayou}
\affiliation{Massachussets Institute of Technology, Cambridge, Massachusetts
02139, USA}

\author{F.~Giuliani}
\affiliation{Istituto Nazionale di Fisica Nucleare, Sezione di Roma, gruppo
collegato Sanit\`a, and Istituto Superiore di Sanit\`a, I-00161 Rome, Italy}

\author{J.~Gomez}
\affiliation{Thomas Jefferson National Accelerator Facility, Newport News,
Virginia 23606, USA}

\author{M.~Gricia}
\affiliation{Istituto Nazionale di Fisica Nucleare, Sezione di Roma, gruppo
collegato Sanit\`a, and Istituto Superiore di Sanit\`a, I-00161 Rome, Italy}

\author{J.O.~Hansen}
\affiliation{Thomas Jefferson National Accelerator Facility, Newport News,
Virginia 23606, USA}

\author{D.~Hayes}
\affiliation{Old Dominion University, Norfolk, Virginia 23508, USA}

\author{D.W.~Higinbotham}
\affiliation{Thomas Jefferson National Accelerator Facility, Newport News,
Virginia 23606, USA}

\author{T.K.~Holmstrom}
\affiliation{College of William and Mary, Williamsburg, Virginia 23187, USA}

\author{C.E.~Hyde}
\affiliation{Old Dominion University, Norfolk, Virginia 23508, USA}
\affiliation{Universit\'{e} Blaise Pascal/IN2P3, F-63177 Aubi\`{e}re, France}

\author{H.F.~Ibrahim}
\affiliation{Old Dominion University, Norfolk, Virginia 23508, USA}
\affiliation{Physics Department, Cairo University, Giza 12613, Egypt}

\author{M.~Iodice}
\affiliation{Istituto Nazionale di Fisica Nucleare, Sezione di Roma Tre,
I-00146 Rome, Italy}

\author{X.~Jiang}
\affiliation{Rutgers, The State University of New Jersey, Piscataway,
New Jersey 08855, USA}

\author{L.J.~Kaufman}
\affiliation{University of Massachussets Amherst, Amherst, Massachusetts
01003, USA}

\author{K.~Kino}
\affiliation{Research Center for Nuclear Physics, Osaka
University, Ibaraki, Osaka 567-0047, Japan}

\author{B.~Kross}
\affiliation{Thomas Jefferson National Accelerator Facility, Newport News,
Virginia 23606, USA}

\author{L.~Lagamba}
\affiliation{Istituto Nazionale di Fisica Nucleare, Sezione di Bari and
University of Bari, I-70126 Bari, Italy}

\author{J.J.~LeRose}
\affiliation{Thomas Jefferson National Accelerator Facility, Newport News,
Virginia 23606, USA}

\author{R.A.~Lindgren}
\affiliation{University of Virginia, Charlottesville, Virginia 22904, USA}

\author{M.~Lucentini}
\affiliation{Istituto Nazionale di Fisica Nucleare, Sezione di Roma, gruppo
collegato Sanit\`a, and Istituto Superiore di Sanit\`a, I-00161 Rome, Italy}

\author{D.J.~Margaziotis}
\affiliation{California State University, Los Angeles, Los Angeles
California 90032, USA}

\author{P.~Markowitz}
\affiliation{Florida International University, Miami, Florida 33199, USA}

\author{Z.E.~Meziani}
\affiliation{Temple University, Philadelphia, Pennsylvania 19122, USA}

\author{K.~McCormick}
\affiliation{Rutgers, The State University of New Jersey, Piscataway,
New Jersey 08855, USA}

\author{R.W.~Michaels}
\affiliation{Thomas Jefferson National Accelerator Facility, Newport News,
Virginia 23606, USA}

\author{D.J.~Millener}
\affiliation{Brookhaven National Laboratory, Upton, New York 11973, USA}

\author{T.~Miyoshi}
\affiliation{Tohoku University, Sendai, 980-8578, Japan}

\author{B.~Moffit}
\affiliation{College of William and Mary, Williamsburg, Virginia 23187, USA}

\author{P.A.~Monaghan}
\affiliation{Massachussets Institute of Technology, Cambridge, Massachusetts
02139, USA}

\author{M.~Moteabbed}
\affiliation{Florida International University, Miami, Florida 33199, USA}

\author{C.~Mu\~noz~Camacho}
\affiliation{CEA Saclay, DAPNIA/SPhN, F-91191 Gif-sur-Yvette, France}

\author{S.~Nanda}
\affiliation{Thomas Jefferson National Accelerator Facility, Newport News,
Virginia 23606, USA}

\author{E.~Nappi}
\affiliation{Istituto Nazionale di Fisica Nucleare, Sezione di Bari and
University of Bari, I-70126 Bari, Italy}

\author{V.V.~Nelyubin}
\affiliation{University of Virginia, Charlottesville, Virginia 22904, USA}

\author{B.E.~Norum}
\affiliation{University of Virginia, Charlottesville, Virginia 22904, USA}

\author{Y.~Okasyasu}
\affiliation{Tohoku University, Sendai, 980-8578, Japan}

\author{K.D.~Paschke}
\affiliation{University of Massachussets Amherst, Amherst, Massachusetts
01003, USA}

\author{C.F.~Perdrisat}
\affiliation{College of William and Mary, Williamsburg, Virginia 23187, USA}

\author{E.~Piasetzky}
\affiliation{School of Physics and Astronomy, Sackler Faculty of Exact
Science, Tel Aviv University, Tel Aviv 69978, Israel}

\author{V.A.~Punjabi}
\affiliation{Norfolk State University, Norfolk, Virginia 23504, USA}

\author{Y.~Qiang}
\affiliation{Massachussets Institute of Technology, Cambridge, Massachusetts
02139, USA}

\author{P.E.~Reimer}
\affiliation{Physics Division, Argonne National Laboratory, Argonne, Illinois 60439, USA}

\author{J.~Reinhold}
\affiliation{Florida International University, Miami, Florida 33199, USA}

\author{B.~Reitz}
\affiliation{Thomas Jefferson National Accelerator Facility, Newport News,
Virginia 23606, USA}

\author{R.E.~Roche}
\affiliation{Florida State University, Tallahassee, Florida 32306, USA}

\author{V.M.~Rodriguez}
\affiliation{University of Houston, Houston, Texas 77204, USA}

\author{A.~Saha}\thanks{Deceased}
\affiliation{Thomas Jefferson National Accelerator Facility, Newport News,
Virginia 23606, USA}

\author{F.~Santavenere}
\affiliation{Istituto Nazionale di Fisica Nucleare, Sezione di Roma, gruppo
collegato Sanit\`a, and Istituto Superiore di Sanit\`a, I-00161 Rome, Italy}

\author{A.J.~Sarty}
\affiliation{St. Mary's University, Halifax, Nova Scotia, Canada}

\author{J.~Segal}
\affiliation{Thomas Jefferson National Accelerator Facility, Newport News,
Virginia 23606, USA}

\author{A.~Shahinyan}
\affiliation{Yerevan Physics Institute, Yerevan, Armenia}

\author{J.~Singh}
\affiliation{University of Virginia, Charlottesville, Virginia 22904, USA}

\author{S.~\v{S}irca}
\affiliation{Dept. of Physics, University of Ljubljana, Slovenia}

\author{R.~Snyder}
\affiliation{University of Virginia, Charlottesville, Virginia 22904, USA}

\author{P.H.~Solvignon}
\affiliation{Temple University, Philadelphia, Pennsylvania 19122, USA}

\author{M.~Sotona}\thanks{Deceased}
\affiliation{Nuclear Physics Institute, \v{R}e\v{z} near Prague, Czech
Republic}

\author{R.~Subedi}
\affiliation{Kent State University, Kent, Ohio 44242, USA}

\author{V.A.~Sulkosky}
\affiliation{College of William and Mary, Williamsburg, Virginia 23187, USA}

\author{T.~Suzuki}
\affiliation{Tohoku University, Sendai, 980-8578, Japan}

\author{H.~Ueno}
\affiliation{Yamagata University, Yamagata 990-8560, Japan}

\author{P.E.~Ulmer}
\affiliation{Old Dominion University, Norfolk, Virginia 23508, USA}

\author{P.~Veneroni}
\affiliation{Istituto Nazionale di Fisica Nucleare, Sezione di Roma, gruppo
collegato Sanit\`a, and Istituto Superiore di Sanit\`a, I-00161 Rome, Italy}

\author{E.~Voutier}
\affiliation{LPSC, Universit\'e Joseph Fourier, CNRS/IN2P3, INPG, F-38026
Grenoble, France}

\author{B.B.~Wojtsekhowski}
\affiliation{Thomas Jefferson National Accelerator Facility, Newport News,
Virginia 23606, USA}

\author{X.~Zheng}
\affiliation{Physics Division, Argonne National Laboratory, Argonne, Illinois 60439, USA}, \affiliation{University of Virginia, Charlottesville, Virginia 22904, USA}

\author{C.~Zorn}
\affiliation{Thomas Jefferson National Accelerator Facility, Newport News,
Virginia 23606, USA}

\collaboration{Jefferson Lab Hall A Collaboration}
\noaffiliation

\date{\today}

\begin{abstract}
\begin{description}
%\begin{itemize} 
\item[Background] In the absence of accurate data on the free 
two-body hyperon-nucleon interaction, the spectra of hypernuclei can provide
information on the details of the effective hyperon-nucleon interaction.
\item[Purpose] To obtain a high-resolution spectrum for the
$^9$Be$(e,e^\prime K^+)$\lamb{9}{Li} reaction.
\item[Method] Electroproduction of the hypernucleus \lamb{9}{Li}
has been studied for the first time with sub-MeV energy resolution in 
Hall A at Jefferson Lab on a $^9$Be  target. In order to increase the
counting rate and to provide unambiguous kaon identification, two 
superconducting septum magnets and a Ring Imaging CHerenkov detector 
(RICH) were added to the Hall A standard equipment.
\item[Results] The cross section to low-lying states of \lamb{9}{Li}
is concentrated within 3 MeV of the ground state and can be fitted
with four peaks. The positions of the doublets 
agree with theory while a disagreement could exist with respect
to the relative strengths of the peaks in the doublets. 
A $\Lambda$ separation energy, $B_{\Lambda}$, of 8.36 $\pm$ 0.08 
(stat.) $\pm$ 0.08 (syst.) MeV was 
measured, in agreement with an earlier experiment.
%There is some disagreement with respect to theory
%both with respect to peak position and cross section.
%\item[Conclusions] Some improvement in reolution and statistics is 
%still needed to resolve the expected low-energy doublet structures in 
%\lamb{9}{Li}.
%\end{itemize}
\end{description}
\end{abstract}


\maketitle

\section{Introduction}
\label{sec:introduction}

Hypernuclei provide a unique laboratory for the investigation of
hadronic many-body systems with strangeness -1 and for the study of 
new aspects of the strong and weak interactions in nuclei. Because 
direct measurements of hyperon-nucleon ($YN$) scattering at low 
energies are prohibited by the short hyperon lifetime, hypernuclear
spectra are the only way to study this interaction. Thus,
a unique opportunity to obtain this information is provided by observing the
fine structure of hypernuclei caused by the specific spin-dependence of
the hyperon-nucleon interaction. Such characteristics are realized in practice
only in $\Lambda$ hypernuclei and are hardly seen in other hypernuclei.
Thus the spectroscopy of $\Lambda$ hypernuclei has a unique value
in strangeness nuclear physics.

 In the past, hypernuclear spectroscopy has been carried out with
limited resolution only by means of hadronic reactions, such as the
strangeness exchange and associated production reactions, that use 
meson beams and in which a neutron is 
converted into a $\Lambda$~\cite{hashtam06}.
More recently, $\gamma$-ray spectroscopy has been used to measure
hypernuclear transition energies. Here, a few-keV energy resolution has
been obtained, which
has allowed precise level assignments and the measurement of doublet
spacings~\cite{tamura08}. However, the method is limited to the bound region
below particle emission thresholds and to bound levels reached following
particle emission.

The experimental knowledge can be enhanced using the $(e,e'K^+)$
electroproduction reaction. This reaction is characterized by a large momentum
transfer to the hypernucleus ($q \gtrsim$ 250 MeV/c) and strong
spin-flip contributions, even at zero $K^+$ production 
angles~\cite{MBSI2010}, 
resulting in the excitation of both natural- and unnatural-parity
states~\cite{iodice07,cusanno09}. In the $(e,e'K^+)$ reaction a proton
is converted into a $\Lambda$ hyperon so that one can produce and
study new hypernuclei, not accessible in the standard reactions.

 Together with experiments in Hall C~\cite{Miyoshi, Yuan, Nakamura},
the E94-107 experiment in Hall A at Jefferson Lab~\cite{proposal}
started a systematic study of high resolution hypernuclear
spectroscopy in the $0p$-shell region of nuclei, such as the
hypernuclei produced in electroproduction on $^9$Be, $^{12}$C, and
$^{16}$O targets. Results on \lam{12}{B} and \lam{16}{N} have
been published~\cite{iodice07,cusanno09}. The results for \lamb{9}{Li},
which was long ago suggested as a good candidate for electroproduction
studies~\cite{sotona94} because of the predicted large splitting of 
the ground-state and second-excited-state doublets are presented in this paper.

\section{Theory}
\label{sec:theory}

 As in the previous experiments~\cite{iodice07,cusanno09},
the data are interpreted using shell-model calculations that
include both $\Lambda$ and $\Sigma$ hyperons in $0s$ states
coupled to $p$-shell core wave functions optimized to fit a wide
range of $p$-shell properties~\cite{millener08,millener12}. 
The $(e,e'K^+)$ reaction is described with distorted-wave impulse 
approximation (DWIA) calculations~\cite{sotona94}  that use
the Saclay-Lyon (SLA) model~\cite{mizutani98} for the elementary
$p(e,e'K^+)\Lambda$ reaction. The SLA model was successfully 
applied in the analysis of electroproduction experiments on $^{12}$C 
and $^{16}$O targets \cite {iodice07, cusanno09}, which suggests 
that this model provides a reasonable prediction for the elementary 
cross section at very small $K^+$ production angles and at the center of mass
energy of this experiment. 

%
% Fig. 1
%
\begin{figure}[t]
\includegraphics[width=8.7cm]{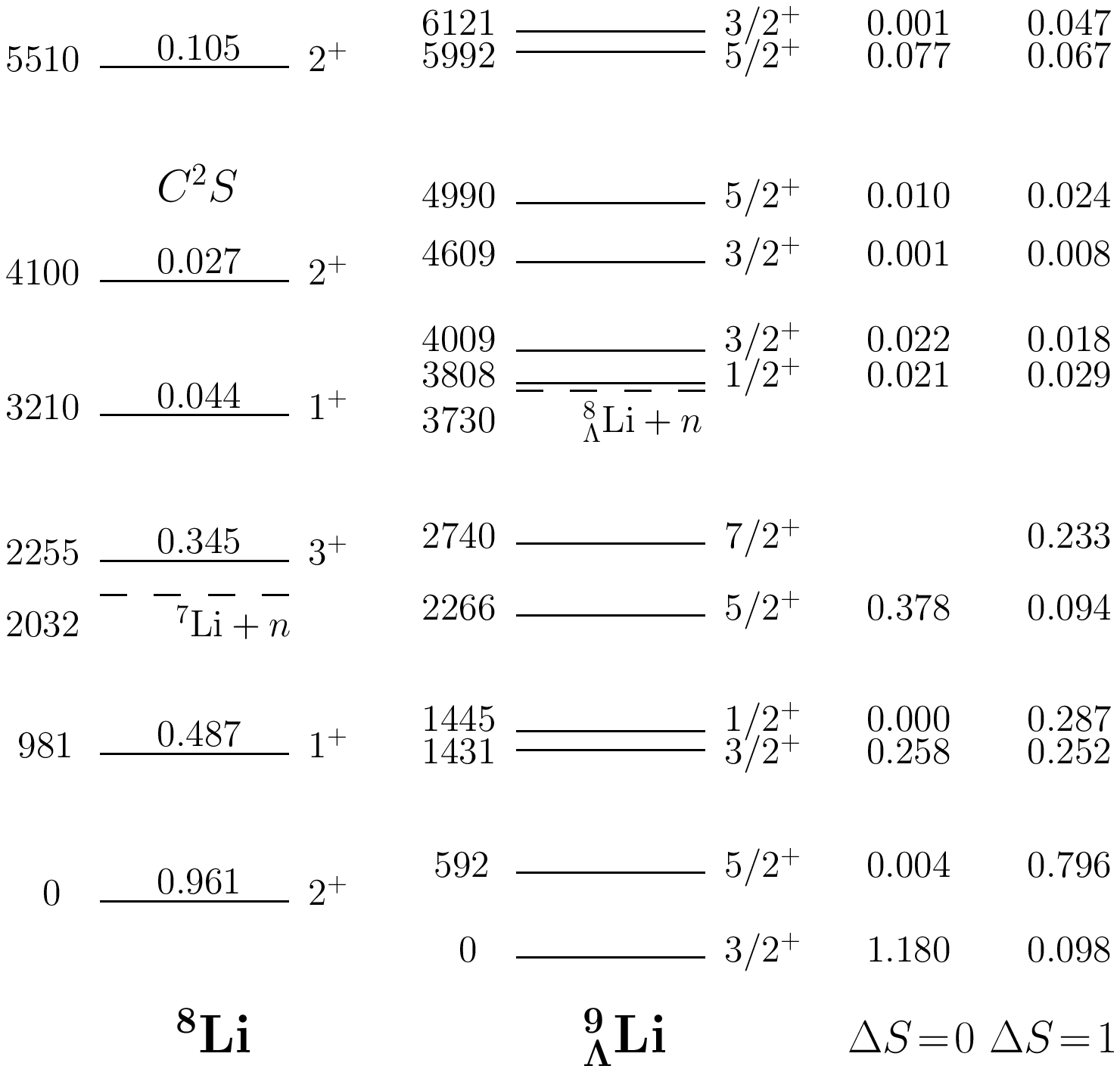}
\caption{The spectrum of \lamb{9}{Li}. The $^8$Li core states are
shown on the left along with the spectroscopic factors for proton
removal from $^9$Be. All excitation energies are in keV. On the right, the
factors  giving the relative population of levels in
purely non-spin-flip ($\Delta S\!=\!0$) and purely spin-flip
($\Delta S\!=\!1$) production reactions on $^9$Be are
given.} \label{fig:lli9}
\end{figure}

In a shell-model approach, one can define five $p_Ns_\Lambda$ two-body
matrix elements for a hypernucleus with an $s$-level $\Lambda$ coupled 
to a $p$-shell nuclear core. These can be put into a one-to-one 
correspondence with the parameters (radial integrals)
$\overline{V}$, $\Delta$, $S_\Lambda$, $S_N$, and $T$ associated with the 
average central, spin-spin, $\Lambda$-spin-orbit, nucleon-spin-orbit, 
and tensor components of the in-medium (effective) $\Lambda N$ 
interaction~\cite{ISS2007}, given by 
\begin{equation}
V_{\Lambda N} = \overline{V} + \Delta \mathbf{s}_N \cdot \mathbf{s}_\Lambda + 
S_\Lambda \mathbf{l}_N \cdot \mathbf{s}_\Lambda + S_N \mathbf{l}_N \cdot 
\mathbf{s}_N + T S_{12},
\label{eq:interaction}
\end{equation}
where $\mathbf{s}$ denotes the spin and $\mathbf{l}$ the angular momentum.
The constant $\overline{V}$ simply contributes $n\overline{V}$ to the
binding energy of every $p^ns_\Lambda$ configuration and therefore does not
affect the spectrum, only the overall binding energy $B_\Lambda$. The
value of $\overline{V} = -1.23$ MeV used is very close to the value that
reproduces the experimental $B_\Lambda$ value (Table 2 of
Ref.~\cite{millener12}). In the weak-coupling limit (quite good because
the $\Lambda N$ interaction for $s_\Lambda$ is a spatial monopole), only
operators that depend on the spin of the $\Lambda$ particle ($\Delta$,
$S_\Lambda$, $T$) contribute to doublet spacings while $S_N$ contributes
to the spacing between doublets.
We use values (in MeV)
\begin{equation}
\Delta= 0.43\quad S_\Lambda =-0.015\quad S_N = -0.39 \quad T =0.03 \; ,
\label{eq:param}
\end{equation}
that fit the spectrum of the five bound levels of
\lamb{7}{Li} determined from \piKg and \Kpig experiments.
The main parameters used for the corresponding $\Lambda N$-$\Sigma N$
interaction are  $\overline{V}'\!=\!1.45$ and $\Delta'\!=\!3.04$ MeV,
making a total of six $YN$ parameters that affect the spectrum.

The calculated spectrum for \lamb{9}{Li} is shown in Fig.~\ref{fig:lli9},
together with the core states for $^8$Li (the first four are known and the
other two are taken from the $p$-shell calculation), while 
Table~\ref{tab:contributions} shows the contributions to the
various level spacings for the three lowest doublets.
The contributions listed in Table~\ref{tab:contributions} do
not add up to exactly the spacings in Fig.~\ref{fig:lli9} because
small contributions from configuration mixing are not included
(see the caption to Table~\ref{tab:contributions}). 
%The values listed 
%in Table~\ref{tab:contributions} together with the differences in core 
%energies do not add up to the spacings in Fig.~\ref{fig:lli9} because 
%contributions from configuration mixing are not included. 
The spectroscopic 
factors ($C^2S$ with $C^2\!=\!2/3$) for proton removal from $^9$Be control 
the population of \lamb{9}{Li} states via electroproduction. The structure 
factors on the right of the figure for pure non-spin-flip and spin-flip
transitions~\cite{millener12} are normalized such that in the
weak-coupling limit ($YN$ interactions turned off) the $\Delta S\!=\!0$
and $\Delta S\!=\!1$ values for a doublet each sum to $C^2S$ for
the core state.

\begin{table}
\caption{Contributions to energy-level spacings (in keV) from
the components of the $\Lambda N$ interaction. The corefficients
of the parameters are determined by numerical differentiation.
The contribution from $\Lambda$-$\Sigma$ coupling is determined
by diagonalizing with the coupling switched on and off. The
difference between the total contribution of 601 keV in the first line
of the table and the 592 keV from diagonalization (see Fig.~\ref{fig:lli9})
is due to small differences in the sum of diagonal core energies
caused by configuration mixing. Such differences are usually only a
few tens of keV.
\label{tab:contributions}}
%Contributions to the energy-level spacings (in keV) 
%from the components of the $\Lambda N$ interaction defined in 
%Eq.~(\ref{eq:interaction}) and the $\Lambda$-$\Sigma$ coupling. 
%\label{tab:contributions}}
\begin{ruledtabular}
\begin{tabular}{lrrrrr}
$J^\pi_i\!-\!J^\pi_f$ & $\Lambda\Sigma$ & $\Delta$ & $S_\Lambda$ & $S_N$
& T  \\
\hline
\\
$\frac{5}{2}^+_1\!-\!\frac{3}{2}^+_1$ & 116 & 531 & $-18$ & $-18$ & $-10$ \\
\\
$\frac{1}{2}^+_1\!-\!\frac{3}{2}^+_2$ & 79 & $-229$ & 13 & 11 & 91 \\
\\
$\frac{7}{2}^+_1\!-\!\frac{5}{2}^+_2$ & 90 & 494 & $-34$ & $-15$ & $-51$ \\
\\
$\frac{3}{2}^+_2\!-\!\frac{3}{2}^+_1$ & 63 & 441 & $-12$ & 56 & $-42$ \\
\\
$\frac{7}{2}^+_1\!-\!\frac{5}{2}^+_1$ & -6 & 8 & $-7$ & $-77$ & $-22$ \\
\end{tabular}
\end{ruledtabular}
\end{table}
\begin{table}
\caption{The $C^2S$ values for proton removal from $^9$Be. The second 
through fourth columns contain the normalized experimental values, 
for the $(d,^3\textrm{He})$ reactions (second and third columns) and 
normalized to the same summed strength (number of p-shell protons in 
$^9$Be) for the $(t,\alpha)$ reaction (fourth column). The values for
one of the other interactions used in hypernuclear calculations are 
listed in the fifth column and the values from the Cohen and Kurath
(6-16)2BME and the (8-16)2BME interactions ~\cite{cohen65} in the 
sixth and seventh columns.
%The values for the (8-16)2BME interaction (seventh column) favor the
%second $1^+$ state over the first, as does the (8-16)POT
%interaction~\cite{cohen67}.
\label{tab:c2s}}
\begin{ruledtabular}
\begin{tabular}{lrrrrrr}
  $J^\pi_i$ & \cite{schwinn75} & \cite{oothoudt77} & \cite{liu88} & fit4
& (6-16)  & (8-16) \\
\hline
\\
 $2^+_1$ & 1.00 & 1.03 & 0.78 & 1.00 & 1.00 & 0.95 \\
\\
 $1^+_1$ & 0.42 & 0.39 & 0.47 & 0.45 & 0.40 & 0.20 \\
\\
 $3^+_1$ & 0.33 & 0.30 & 0.51 & 0.36 & 0.35 & 0.33 \\
\\
 $1^+_2$ &      &      &      & 0.04 & 0.06 & 0.24 \\
\end{tabular}
\end{ruledtabular}
\end{table}

 The $C^2S$ values for $^9\textrm{Be}\to {^8}$Li in Fig.\ref{fig:lli9}
are in good agreement with the values from $(d,^3\textrm{He})$
studies~\cite{schwinn75,oothoudt77} (see Table~\ref{tab:c2s}).
From a $(t,\alpha)$ study~\cite{liu88}, larger relative values were
extracted for the excited $1^+$ and $3^+$ states of $^8$Li. 
The $C^2S$ values for the interaction used in the present work are listed 
in Fig. \ref{fig:lli9}.
The values for all the $p$-shell interactions derived in connection with these
hypernuclear studies are similar and in agreement with those
 for the Cohen and Kurath (6-16)2BME interaction~\cite{cohen67}.
The values for the other two Cohen and Kurath interactions put more
strength in the second $1^+$ state than in the first (as noted in
Ref.~\cite{cohen67} for the (8-16)POT interaction). The reason for
this is that the $1^+_1$ states are rather purely $L\!=\!1$, $S\!=\!1$,
rather than with strongly mixed $S\!=\!0$ and $S\!=\!1$ components,
as happens for the other interactions. Strength for the $1^+_2$ state
would be immediately noticeable in electroproduction because the
\lamb{9}{Li} states based on the $1^+_2$ state lie close to the neutron
threshold at 3.73 MeV and should therefore be narrow. 

 The states in the first-excited doublet are predicted to be nearly
degenerate, in part because of the contributions from $\Lambda$-$\Sigma$
coupling (see Table~\ref{tab:contributions}). In addition, the $3/2^+_2$
state contains a 3.5\% admixture of a $\Lambda$ coupled to the $^8$Li
ground state which lowers the $3/2^+_2$ state by another $\approx 35$ keV.
Thus, if the shell-model predictions are reasonable,
five peaks should, in principle, be resolved in \lamb{9}{Li} below the 
particle-decay threshold by an electroproduction
experiment with good energy resolution.

\section{Experiment}
\label{sec:experiment}

%
% Fig. 2
%
\begin{figure*}
\includegraphics*[width=17.5cm]{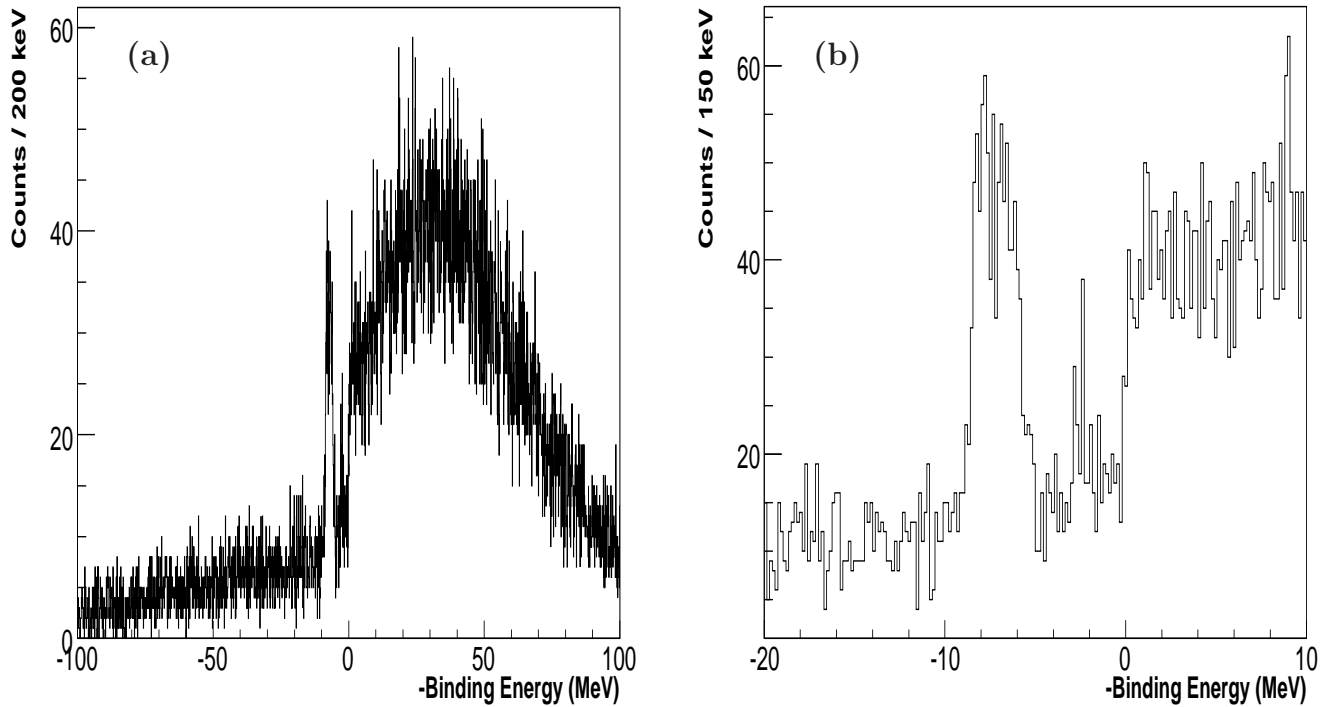}
\caption{The binding-energy spectrum obtained after kaon selection with 
aerogel detectors and RICH in (a) the whole energy range and (b) restricted 
to the region of interest.}
\label{be1}
\end{figure*}

Hall A at JLab is well suited to perform $(e,e'K^+)$ experiments.
Scattered electrons are detected in the High Resolution Spectrometer
(HRS) electron arm while coincident kaons are detected in the HRS hadron
arm~\cite{nimhalla}. The disadvantage of smaller electromagnetic cross
sections is compensated for by the high current and high duty
cycle properties of the beam.
Throughout the experiment, the same equipment has been used in very
similar kinematical conditions on C, Be, and H$_2$O targets.
The use of a pair of septum magnets permitted particle detection at very
forward angles~\cite{septum} and a Ring Imaging CHerenkov (RICH)
detector~\cite{rich2004,pylosfg,pylosfc,rich_algorithm} has been used 
in the hadron arm to
provide an unambiguous identification of kaons when combined with
the standard particle identification apparatus of Hall A, based on aerogel
Cherenkov detectors~\cite{perrino,lagamba,marrone}. 
%Details and motivations
%for the specific choices can be found in~\cite{iodice07}.
In the present experiment a 92.5~mg/cm$^2$ solid $^{9}$Be target with a
beam current of $\sim$100~$\mu$A was used at a beam energy of 3775 MeV.
Both HRSs were physically positioned at an angle of 12.5$^{\circ}$, but 
the pair of septum magnets yielded an effective angle for both the scattered 
electron and the hadron detection of $\sim6 ^{\circ}$.

Fig.~\ref{be1} shows the observed binding-energy spectrum
of \lamb{9}{Li}. The broad peak centered at a small positive binding energy
corresponds to the \lamb{9}{Li} states in Fig.~\ref{fig:lli9} corresponding to 
the lowest three
states of $^8$Li. The rise in cross section starting at 0 MeV
corresponds to states with the $\Lambda$ in a $p$ orbit and, because
these states are unbound, the states are broad and no structure is observed.
As in Ref.~\cite{iodice07,cusanno09}, the background was determined from the 
binding energy spectrum obtained with a coincidence time shifted with respect 
to the coincidence time between secondary electrons and produced kaons and was 
rather flat for values of binding energy ranging from 15 MeV
to 0 MeV. Its value was calculated as the average of the counts in 
the range 9.95~MeV $\leq$ Binding energy $\leq$ 18.35~MeV. 

 For the calculation of the absolute cross section, we computed the
following quantities: detector efficiencies, detector dead time,
detector phase space, kaon survival in HRS, integrated luminosity. 
The calculation of efficiencies for the standard HRS package are 
well established and implemented in the Hall A analysis software. 
Therefore, those procedures were used for that purpose. 
For the RICH and aerogel Cherenkov detectors, we used one detector to 
determine the efficiency of the other one in the following way: 
we selected a pure sample of kaons by means of aerogel detectors and we 
measured the fraction of those kaons detected by the RICH and vice versa.
The detector dead time was measured by the Hall A data acquisition system.
The detector phase space was calculated using the SIMC code~\cite{SIMC}.
Kaon survival is calculated considering the average path length inside 
the HRS arm. The integrated luminosity was calculated by means of beam 
current monitor devices. Then, the absolute cross section $\sigma$
was computed according to 
\begin{equation}
\sigma = \frac{Counts}{Ksur \cdot Eff \cdot Luminosity \cdot PhaseSpace
\cdot Livetime}\, ,
\label{eq:3}
\end{equation}
%Cross section = Counts/Ksurv/TotEff/Luminosity/phaceSpace
where $Counts$ is the event number in the experiment,
$Eff$ is the global detector efficiency, 
$ Livetime$ is 1-detector dead time,
$ PhaseSpace$ is the detector phase space, 
$Ksur$ is the kaon survival in the HRS, 
and $ Luminosity$ is the integrated luminosity.

\begin{figure}
\centering
\includegraphics[width=8.5cm]{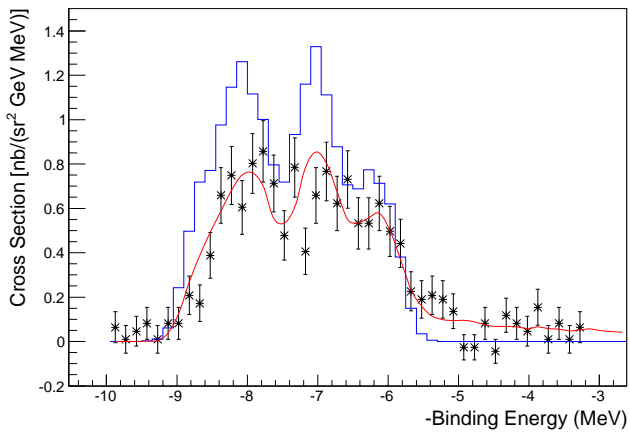}
\caption{(Color online) The \lamb{9}{Li} differential cross section
as a function of the binding energy.
Experimental points vs. Monte Carlo results (red curve) and vs.
 Monte Carlo results with radiative effects turned off (blue histogram).}
\label{simulbe}
\end{figure}

Fig.~\ref{simulbe} shows the background-subtracted
experimental binding-energy spectrum, together with Monte Carlo 
simulations\cite{SIMC} (red curve) and the same simulations 
with the radiative effects turned off (blue curve). 
The error bars in the data are statistical. The simulations used the five
peak positions and widths listed as configuration $\alpha$ in 
Table~\ref{Peak_Configurations}. 
The red curve fits the experimental data well with a  corresponding
$\chi^2$/ndf value of 36.69/35. Several other peak configurations,
with different numbers, heights, positions and widths of the peaks, have been 
found to reproduce the red curve. All of those are also expected to
generate the same spectrum (the blue curve of Fig.~\ref{simulbe}) when
radiative corrections are turned off, since
radiative corrections are independent of the assumptions regarding the
number and type of the peaks that build up the experimental spectrum.
In practice, the simulated data do not overlap
perfectly with the experimental data, which produces
small systematic errors on the radiatively corrected spectrum.
%
% Fig. 3
%
%
% Fig.4
%
\begin{figure}
\centering
\includegraphics[width=8.5cm]{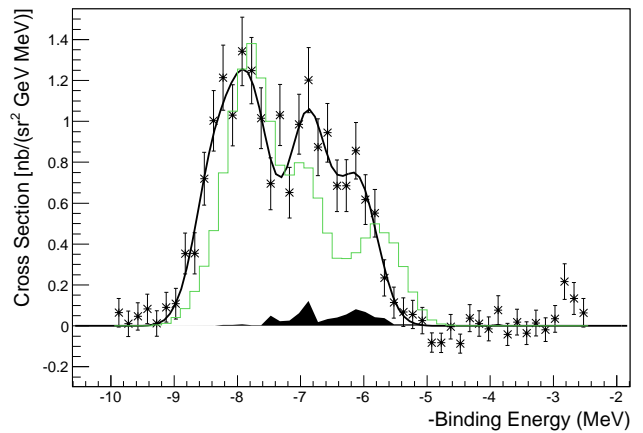}
\caption{(Color online) The radiatively unfolded experimental spectrum compared
to a theoretical prediction (thin green line). The solid black line 
represents a fit to the data with four Gaussians of a common width. The 
theoretical curve was calculated with the width extracted from the fit 
(FWHM = 730 keV).}
\label{databe}
\end{figure}

The unfolding for radiative corrections has been done 
bin-by-bin. The content of each bin of the radiatively corrected
spectrum was obtained by multiplying the corresponding bin of the
experimental spectrum by the ratio of the blue and red curves of 
Fig.~\ref{simulbe} for that bin. In order to avoid possible removals of 
background enhancements or to artificially null the spectrum in the 
regions where the blue curve is zero, the ratio between the blue and 
red curves of Fig.~\ref{simulbe} was performed after summing the background 
to each of them. The background value was then subtracted from the result of 
the product of the ratio with the corresponding bin. The result is shown 
in Fig.~\ref{databe} which presents the radiatively unfolded
experimental data (points with statistical errors) compared to a~theoretical
prediction (thin green line). The band at the bottom of the histogram 
represents the systematic errors in the radiative unfolding. The theoretical 
histogram was obtained
using the procedure described in section \ref{sec:theory} assuming an energy 
resolution of 730~keV (FWHM). Once radiative
corrections have been applied, the binding-energy spectrum resolution
is small enough to clearly show a three-peak structure in the spectrum. A more
detailed description of the procedure employed to determine the radiatively
unfolded spectrum of  Fig.~\ref{databe} is given 
in Appendix~\ref{Appendix A}.

%
% Table III
%
\begin{table*}
\caption{Excitation energies, widths, and cross sections obtained by
fitting the $^{9}$Be$(e,e'K^+)$\lam{9}{Li} spectrum (first three columns),
compared with theoretical predictions (last four columns). The last
column gives the summed cross sections for the three doublets, to be compared 
with the experimental results in the third column.
\label{tab:results}}
\begin{ruledtabular}
\begin{tabular}{cccccccc}
\multicolumn{3}{c}{Experimental data}  &
\multicolumn{4}{c}{Theoretical predictions} \\
$E_x$ &  Width (FWHM) &  Cross section & $E_x$  & $J^\pi$ & Cross section
&  Cross section \\
 (MeV)  &  (MeV) & (nb/(sr$^2$ GeV)) & (MeV) &  &
 (nb/(sr$^2$ GeV)) & Sum \\
 \hline
 \phantom{1}
 0.00 $\pm$ 0.08 & 0.73 $\pm$ 0.06 & 0.59 $\pm$ 0.15 & 0.00
                                   & $3/2^+$ & 0.18 & 1.22 \\
 \phantom{1}
 0.57 $\pm$ 0.12 & 0.73 $\pm$ 0.06 & 0.83 $\pm$ 0.13 & 0.59 
                                   & $5/2^+$ & 1.04 & \\
 & & & & & \\
\phantom{1}1.47 $\pm$ 0.09 & 0.73 $\pm$ 0.06 & 0.79 $\pm$ 0.07 & 1.43 &
$3/2^+$ & 0.29 & 0.59\\
   &  &  & 1.45 & $1/2^+$   &0.30 & \\
 & & & & \\
\phantom{1}2.27 $\pm$ 0.09   & 0.73 $\pm$ 0.06 & 0.54 $\pm$ 0.06 & 2.27
& $5/2^+$ & 0.17 & 0.48 \\
   &  &  & 2.74 & $7/2^+$ & 0.31 & \\
 & & & & & \\
\end{tabular}
\end{ruledtabular}
\end{table*}

\section{Results}
\label{sec:results}

When analyzing the  experimental spectrum in Fig.~\ref{databe} 
one has to consider that, as explained in section~\ref{sec:theory},
the spectrum is made up by doublets and hence that each of 
the three peaks that appears in it is actually produced by the 
convolution of two ``elementary'' peaks. Because the peaks of the 
spectrum are radiatively corrected, we assumed that all the elementary 
peaks were well described by Gaussian distributions. Considering the 
energy resolution to be constant over the whole 
spectrum range, we assumed in addition that the standard deviations
of these Gaussians were equal.
Although six Gaussian elementary peaks are expected, the possible
existence of nearly degenerate doublets, or of doublets where one peak 
overwhelmingly dominates on the other, could reduce the number of elementary
Gaussian peaks needed for the fit procedure following the Occam razor 
principle. The experimental spectrum in Fig.~\ref{databe} was fitted 
in order to determine the positions, the heights, and the common standard 
deviation of the elementary peaks. 
The best fit was obtained with four Gaussian elementary peaks  
with a $\chi^2$/n.d.f. value of 41.82/41. 
The energy resolution extracted from the fit, 730 keV (FWHM), is consistent 
with the value obtained in our previous analysis \cite{Cusanno2010} and
is in agreement with the measurements on  \lam{12}{B} \cite{iodice07} 
and \lam{16}{N} \cite{cusanno09}. The excitation energies ($E_x$) 
and cross sections extracted from the four-peak
fit are reported in Table~\ref{tab:results} where they are
compared with the results calculated using the procedure described in 
section \ref{sec:theory} for the six lowest states shown in 
Fig.~\ref{fig:lli9}. A fit with five Gaussian peaks produced the same result
as shown in Table~\ref{tab:results} with a $\chi^2$/ndf value of 41.82/39
and a common FWHM for the peaks of 730 keV. The first three peaks 
had the same heights and positions of the corresponding peaks in
Table~\ref{tab:results} while the fourth and fifth peaks had
equal positions, coincident with the binding energy of the fourth peak in   
Table~\ref{tab:results} and heights whose sum was equal to 
the cross section value of the fourth peak in 
Table~\ref{tab:results}. The result of the five-peak fit showed hence
that only the ground-state doublet splitting could be detectable 
with the energy resolution of the experiment.
A fit with three peaks also produced a result consistent with 
Table~\ref{tab:results}, with a $\chi^2$/ndf value of 47.52/43 and
a common peak resolution extracted from the fit of 970 keV (FWHM). 
The first peak's strength and position were equal (within one standard 
deviation) to the sum of the strengths and to the baricenter of the positions
in the binding-energy spectrum of the first two states of 
Table~\ref{tab:results} respectively. The other two peaks had strengths and 
positions  equal (within one standard deviation) to the third and fourth peaks 
in Table ~\ref{tab:results} respectively.
Fig.~\ref{databe} and Table ~\ref{tab:results} show 
that the observed peak positions agree quite well with the predictions of
the standard model for $p$-shell hypernuclei.
The first multiplet can be decomposed into two peaks with 
a separation of $570 \pm 120$ keV that 
corresponds very well with the theoretical value of 590 keV.
On the other hand, there is a systematic disagreement for the 
multiplet cross sections. 
In the first multiplet the 0.59 MeV (5/2$^+$) peak does not dominate as 
theoretically predicted (see Table ~\ref{tab:results}).
The second and third multiplets are each observed as a single peak.
This is probably due to the very close excitation 
energies of their two constituents (see Table ~\ref{tab:results}), 
although for the third multiplet it might be due to the fact 
that the strength of the 2.27 MeV (5/2$^+$) peak dominates over that 
of the other state.

In terms of the cross section, the spin-spin
interaction ($\Delta$) tends to deplete the spin-flip strength to
the ground-state doublet and increase the non-spin-flip strength
(see Fig.~\ref{fig:lli9}). The full reaction calculations include a
number of spin-flip and non-spin-flip amplitudes, making the
cross sections sensitive to the choice of the elementary reaction
model. The SLA model was selected from the various isobar models 
because it gives the best results for the cross section. Spin-flip 
amplitudes are dominant in the SLA model which favors states in
Fig.~\ref{fig:lli9} with large $\Delta S\!=\!1$ structure amplitudes.
It is then clear that a model with larger non-spin-flip amplitudes
might increase the relative cross sections for the $3/2^+_1$ and
$5/2^+_2$ states and provide better agreement with the results
of the experimental analysis. 
%The theoretical 
%result for the second member of the ground-state doublet (5/2$^+$) 
%agrees quite well with the experimental value  
%but predictions for the ground state and the excited-state doublets 
%are about two times smaller (see columns 3 and 8 in Tab.III).   
The cross section depends very much on the proton
removal spectroscopic factors for $^9$Be but, as is evident
from Table~\ref{tab:c2s}, theory agrees very well with the relative
$C^2S$ values derived from the analysis of two $(d,^3\textrm{He})$
studies, a reaction that has proven to be very reliable for such a
comparison.

From the binding-energy spectrum of Fig.~\ref{databe}, a $\Lambda$ separation
energy $B_{\Lambda}$ of 8.36 $\pm$ 0.08 (stat.) $\pm$ 0.08 (syst.) MeV 
was obtained. 
This value agrees very well with the value $8.50\pm 0.12$ MeV from
emulsion data \cite{Carbon_Ground_State}.
To determine  this value the missing-mass scale needed to be calibrated
because of uncertainties in the kinematical variables such
as the primary electron energy and the central momenta and the central 
scattering angles of the scattered electrons and 
the produced kaons. 
For this calibration we took advantage of the fact that the experiment was 
performed just after the determination of the \lam{12}{B}  excitation spectrum 
\cite{iodice07} that used the same experimental settings. Thus, the 
kinematical variables of the  present experiment were determined, reproducing
the binding energy of the \lam{12}{B} ground state at 11.37 $\pm$ 0.06 MeV 
\cite{Carbon_Ground_State}.
A more detailed description of this missing-mass scale calibration is given 
in Appendix~\ref{Appendix B}.

\section{summary}
\label{sec:summary}

 A high-quality \lamb{9}{Li} hypernuclear spectrum
has been obtained for the first time with sub-MeV energy resolution.
The measured cross sections and the excitation energies of the 
doublets are in a good agreement with the values predicted using the SLA model 
and simple shell-model wave functions while a disagreement could exist with 
respect to the relative strengths of the states making up the first multiplet. 
As noted in the Sec.~\ref{sec:results}, an elementary model for the 
$(e,e'K^+)$ reaction with a different balance of spin-flip and non-spin-flip
amplitudes might help to resolve this disagreement.
A $\Lambda$ separation energy $B_{\Lambda}$ of 8.36 $\pm$ 0.08 
(stat.) $\pm$ 0.08 (syst.) MeV was 
obtained, in good agreement with the emulsion value.

\appendix

\section{Radiative corrections}
\label{Appendix A}

The procedure of unfolding radiative effects from an experimental spectrum 
does not depend on the choice 
of the peak structure used to fit the spectrum itself, providing that the fit 
describes the data reasonably. This property is very useful when the peak 
structure underlying an experimental spectrum is 
unknown as in  Fig.~\ref{simulbe}, where several peak structures fit 
the experimental spectrum quite well and it is not 
obvious which of these structures is ``the right one''.
To demonstrate the independence of radiative corrections 
from the energy spectrum structure, we define $Exp(E)$ as the function that 
describes the experimental spectrum. $Exp(E) \cdot dE$ is proportional to the 
number of events whose corresponding energy is in the interval $E \pm dE$. 
We define $S(E')$ as the function that describes the experimental 
spectrum in the absence of radiative effects. Lastly, we define $R(E'-E)$ 
as the probability that an event whose corresponding
energy in the absence of radiative effects would have been $E'$ has, 
because of the radiative effects, an energy equal to $E$.
$Exp(E)$, $S(E')$ and $R(E-E')$ are related by 
\begin{equation}
Exp(E) \cdot dE = dE \cdot \int_{}^{}{dE' \cdot  R(E'-E) \cdot S(E')}
\label{eq:4}
\end{equation}
For the sake of simplicity, we suppose in the following that 
$S(E')$ is equal to a sum of Gaussian peaks
\begin{equation}
S(E') = \sum_{k=1}^N{A_k \cdot e^{-\frac{(E' - E_k)^2}{2 \cdot \sigma_k^2}}}
\ ,
\label{eq:5}
\end{equation}
where $A_k$, $E_k$ and $\sigma_k$ are the amplitude, central value and
standard deviation of the kth peak, respectively.  

Let us assume two different peak configurations $\alpha$ and
$\beta$, with $N$ and $M$ peaks, respectively, that produce two
functions $S^{\alpha}(E')$ and $S^{\beta}(E')$ that are equal within 
the statistical error 
\begin{equation}
S^{\alpha}(E') = 
\sum_{k=1}^N{A_k  e^{-\frac{(E' - E_k)^2}{2 \cdot \sigma_k^2}}} \approx
\sum_{l=1}^M{A_l  e^{-\frac{(E' - E_l)^2}{2 \cdot \sigma_l^2}}} =
S^{\beta}(E')
\label{eq:6}
\end{equation}
This implies that for every value of $E'$,
$S^{\alpha}(E')$ and $S^{\beta}(E')$ have statistically compatible values 
and the $\chi^2$ test
\begin{equation}
\chi^2 = \sum_{j}{\frac{(S_j^{\alpha}-S_j^{\beta})^2}{S_j^{\alpha}}}\; ,
\label{eq:7}
\end{equation}
with 
\begin{equation}
S^{\alpha,\beta}_j =\int_{E'_{j-1}}^{E'_j}{dE' \cdot S^{\alpha,\beta}(E')}\; ,
\label{eq:8}
\end{equation}
is acceptable within our confidence level. In Eq.~(\ref{eq:8}), 
$[E_{j-1}; Ej]$ is the $jth$ interval that the energy spectrum is divided into.

\begin{figure}
\includegraphics[width=8.5cm]{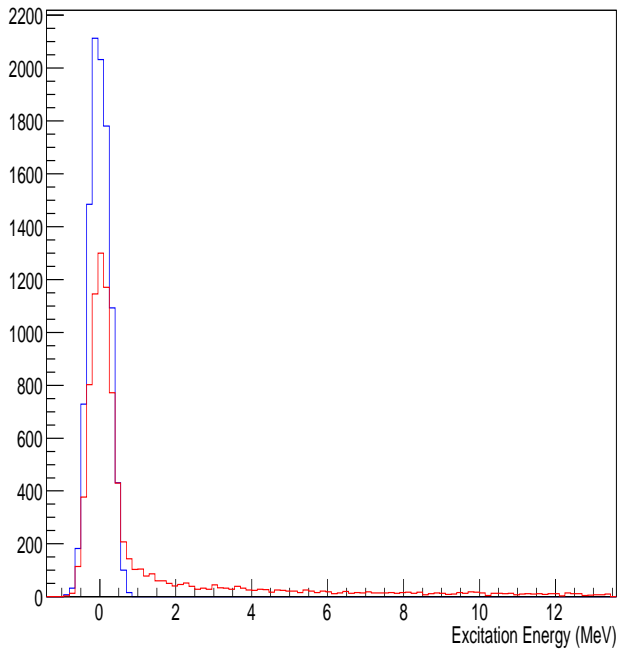} 
\caption{(Color online) One peak of the  excitation energy spectrum of the 
hypernucleus \lamb{9}{Li} obtained through  the reaction 
$^9$Be$(e,e'K^+)$\lamb{9}{Li} as predicted by the Monte Carlo SIMC when 
including all effects (red curve) and ``turning off'' the radiative effects 
(blue curve). Arbitrary units. The position of the peak has  been made 
coincident with the ground state.}
\label{Single_Peaks}
\end{figure}

It is obvious from Eq.~(\ref{eq:4}) that if $S^{\alpha}(E') = S^{\beta}(E')$ 
the two peak configurations $\alpha$ and $\beta$ will produce the same 
experimental spectrum, that is $Exp^{\alpha}(E) = Exp^{\beta}(E)$. 

The reverse is also true: if two peak configurations $\alpha$ and $\beta$
produce two statistically compatible spectra
($Exp^{\alpha}(E) = Exp^{\beta}(E)$) then $S^{\alpha}(E') = S^{\beta}(E')$.
In fact, defining
\begin{equation}
Exp_i = \int_{E_{i-1}}^{E_i}{dE \cdot Exp(E)}\; ,
\label{eq:9}
\end{equation}
and
\begin{equation}
R_{ij} = \int_{E_{i-1}}^{E_i}{dE \cdot R(E'-E)}\: ,  
(E' \in [E'_{j-1};E'_j]) 
\label{eq:10}
\end{equation} 
we have from Eq.~(\ref{eq:4})
\begin{equation}
\int_{E_{i-1}}^{E_i}{dE \cdot Exp(E)} = 
\int_{}^{}{dE' \cdot  S(E') \int_{E_{i-1}}^{E_i}{dE \cdot  R(E'-E)}}
\label{eq:11}
\end{equation}
Eq.~(\ref{eq:11}) means that
\begin{equation}
Exp_i  = \sum_{j}^{}{R_{ij} \cdot S_j}
\label{eq:12}
\end{equation}
or, defining the arrays 
$\overrightarrow{Exp} \equiv \{Exp_1, Exp_2, ... Exp_i, ...\}$ and 
$\overrightarrow S \equiv \{S_1, S_2, ... S_j, ...\}$, and the matrix
$R \equiv \{R_{11}, R_{12}, ..., R_{ij}, ...\}$ 
\begin{equation}
\overrightarrow{Exp}  = R \cdot \overrightarrow S
\label{eq:13}
\end{equation}
Defining at last $R^{-1}$ as the inverse of the matrix $R$, we have 
\begin{equation}
\overrightarrow S  = R^{-1} \cdot \overrightarrow{Exp}
\label{eq:14}
\end{equation}
From Eq.~(\ref{eq:14}) it follows that if $Exp^{\alpha}(E) = 
Exp^{\beta}(E)$ then $S^{\alpha}(E') = S^{\beta}(E')$. In fact,
\begin{table}
\caption{Columns 2 and 3: peak positions and relative amplitudes of five 
configurations $\alpha$, $\beta$, $\gamma$, $\delta$ and $\epsilon$ for which
the Monte Carlo SIMC predicts a \lamb{9}{Li} excitation
energy spectrum that fits the experimental data. Column 4: the $\chi^2$ test 
values calculated through Eq.~(\ref{eq:16}) for these configurations.}
\label{Peak_Configurations}
\begin{ruledtabular}
\begin{tabular}{cccc}
 Configuration  & Peak Positions & Peak Amplitudes & $\chi^2$  \\ 
   & MeV & Arbitrary units & 35 ndf \\
\hline
            & 0.00    & 2.23  &          \\
            & 0.64    & 3.54  &          \\
   $\alpha$ & 1.32    & 1.90  & 36.685   \\
            & 1.71    & 2.61  &          \\
            & 2.35    & 2.33  &          \\[5pt]
            & 0.00    & 2.08  &          \\
            & 0.58    & 3.48  &          \\
   $\beta$  & 1.54    & 3.38  & 38.247   \\
            & 2.37    & 2.10  &          \\[5pt]
            & 0.00    & 2.34  &          \\
            & 0.54    & 3.88  &          \\
   $\gamma$ & 1.49    & 3.78  & 46.088   \\
            & 2.36    & 3.28  &          \\[5pt]
            & 0.00    & 1.86  &          \\
            & 0.54    & 3.08  &          \\
   $\delta$ & 1.49    & 3.00  & 39.068   \\
            & 2.36    & 2.06  &          \\[5pt]
            & 0.00    & 1.85  &          \\
            & 0.65    & 3.09  &          \\
 $\epsilon$ & 1.43    & 3.00  & 39.000   \\
            & 2.39    & 2.06  &          \\
\end{tabular}
\end{ruledtabular}
\end{table}
\begin{eqnarray}
Exp^{\alpha} = Exp^{\beta} \Rightarrow
0 & = & R^{-1} \cdot (\overrightarrow {Exp^{\alpha}} - \overrightarrow 
{Exp^{\beta}}) \nonumber \\  
& = & \overrightarrow {S^{\alpha}} - \overrightarrow {S^{\beta}} \Rightarrow 
S^{\alpha} = S^{\beta}
\label{eq:15}
\end{eqnarray}
means that the spectrum with the radiative effects subtracted ($S(E')$)
does not depend on the peak configurations $\alpha$, $\beta$, $\cdots$ as long 
as all the configurations considered $Exp^{\alpha}$, $Exp^{\beta}$, $\cdots$
fit the experimental spectrum with no radiative effects applied.   
It has to be noted that only two assumptions were made in deriving the 
conclusion quoted above

\begin{figure*}
\begin{center}
\includegraphics*[width=17.5cm]{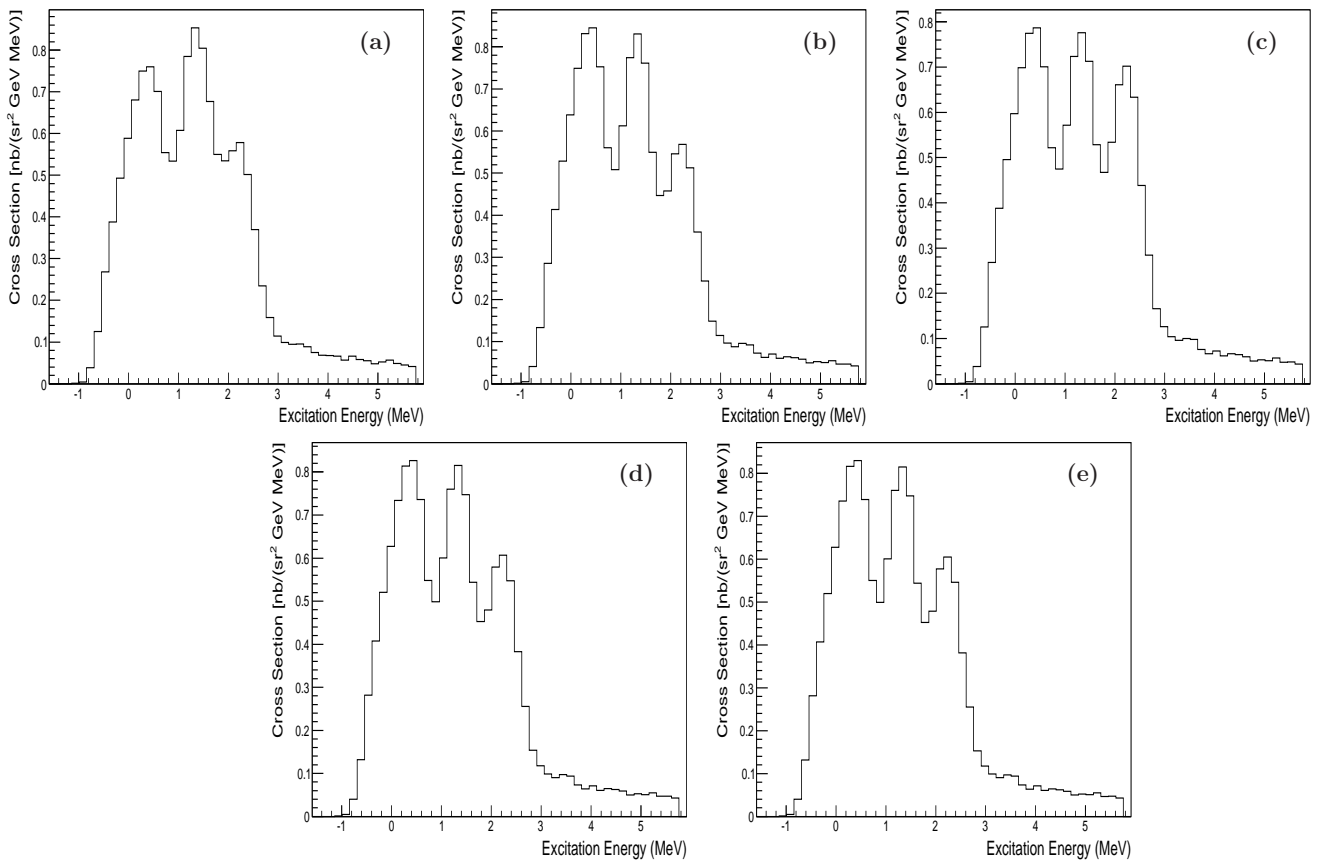}
\caption{\lamb{9}{Li} excitation energy spectra as predicted by the Monte 
Carlo SIMC for the peak configurations $\alpha$, $\beta$, $\gamma$, $\delta$ 
and $\epsilon$ quoted in Table \ref{Peak_Configurations}
(Panels (a), (b), (c), (d) and (e), respectively).} 
\label{Monte_Carlo_Spectra With_Radiative_Effects}
\end{center}
\end{figure*}

\begin{itemize}
\item the single intervals $[E_{j-1}; E_j]$ are so small that $R_{ij}$
defined by Eq.~(\ref{eq:10}) is constant in it.
\item The matrix $R$ is invertible. This is usually the case considering that
usually $R_{ii} \neq 0$ and $R_{ji} = 0$ if $R_{ij} \neq 0$ and 
$i \neq j$
\end{itemize}

\begin{figure*}
\begin{center}
\includegraphics*[width=17.5cm]{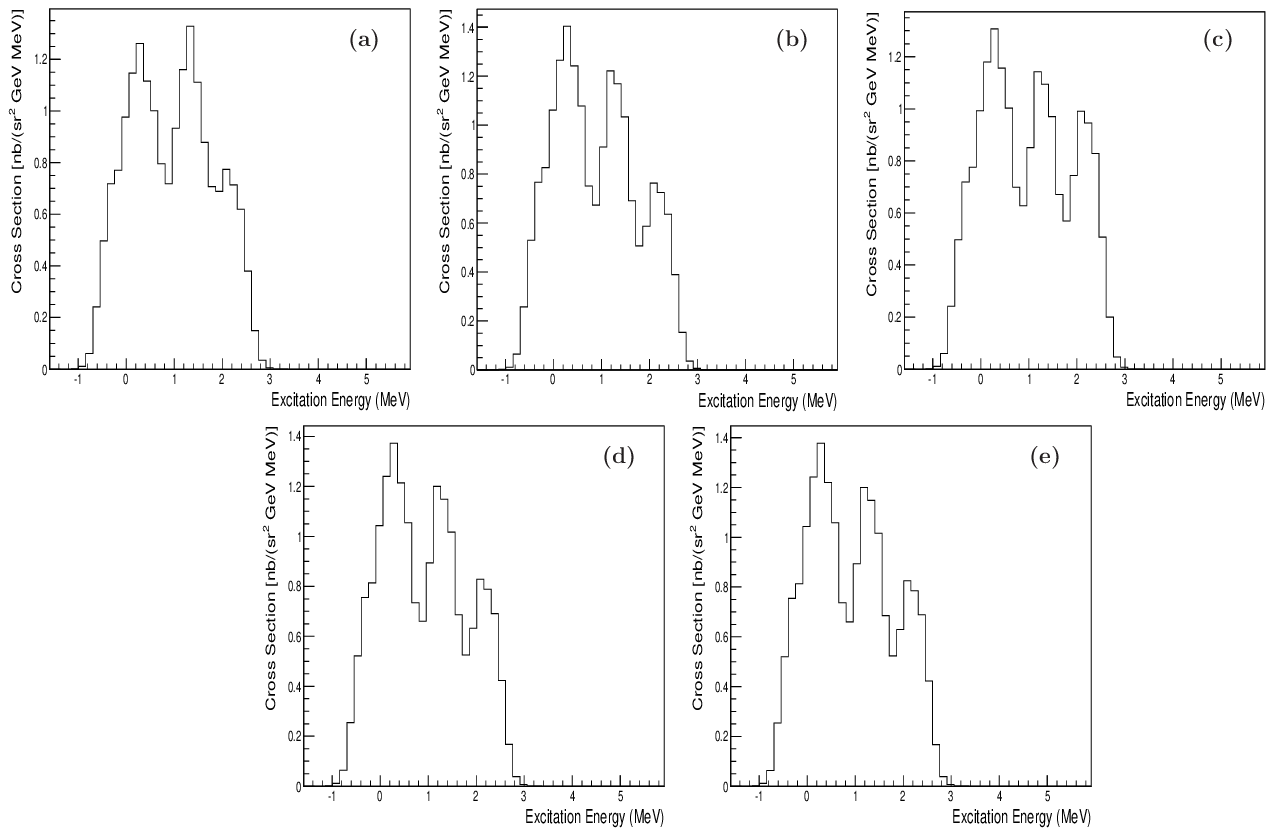}
\caption{\lamb{9}{Li} excitation energy spectra as predicted by the Monte 
Carlo  SIMC for the peak configurations $\alpha$, $\beta$, $\gamma$, $\delta$ 
and $\epsilon$ quoted in Table  \ref{Peak_Configurations},  
when the radiative effects are ``turned off'' 
(Panels (a), (b), (c), (d) and (e), respectively).}
\label{Monte_Carlo_Spectra Without_Radiative_Effects}
\end{center}
\end{figure*}

\begin{figure*}
\includegraphics*[width=18.0cm]{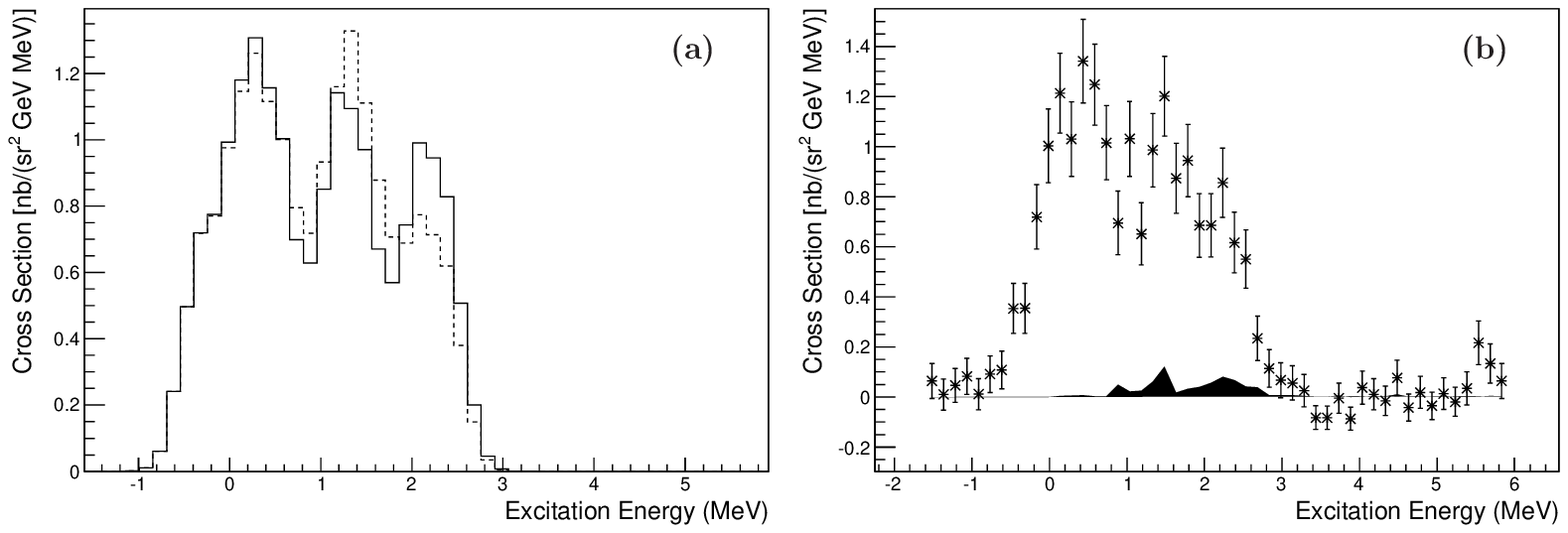}
\caption{(a) the spectrum of the configuration $\alpha$ (dashed line) and of 
the configutration $\gamma$ (continuous line) as predicted by the Monte Carlo
SIMC when the radiative effects are ``turned off''. (b) the statistical errors 
(error bars) and the systematic errors (full band) as a function of the 
excitation energy. The systematic error was defined as the difference 
between the dashed line and the continuous line of panel (a), see text for 
details.}
\label{Histogram_Comparisons}
\end{figure*}

To determine the spectrum with the radiative effects ``turned off'' (blue
curve of Fig.~\ref{simulbe}) the Monte Carlo SIMC was employed~\cite{SIMC}.
The red curve of Fig.~\ref{Single_Peaks} shows a single peak of the 
\lamb{9}{Li} excitation energy spectrum as predicted  by SIMC for the reaction 
$^9$Be$(e,e'K^+)$\lamb{9}{Li} when using the E94-107 experimental apparatus 
(position and amplitude of the peak are arbitrary). The blue curve of 
Fig.~\ref{Single_Peaks} shows the same peak when the radiative effects in the 
Monte Carlo SIMC are ``turned off''. Several peak configurations, 
made up by a number of peaks like the one of Fig.~\ref{Single_Peaks} red curve, 
fit the  experimental \lamb{9}{Li} excitation energy  spectrum after 
being normalized to it. Table~\ref{Peak_Configurations} quotes five of 
them. Their corresponding excitation energy spectra (normalized to 
the experimental data) are shown in 
Fig~\ref{Monte_Carlo_Spectra With_Radiative_Effects}.
For each configuration, Table \ref{Peak_Configurations} quotes the position 
and relative amplitude of the peaks (here and in the following 
the amplitude of a peak is defined as the integral of the peak over the whole
energy spectrum) and the value of the $\chi^2$ test,
\begin{equation}
\chi^2  = \sum_{i}^{}{\frac{(C^{conf}_i-Exp_i)^2}{Exp_i}}\: ,
\label{eq:16}
\end{equation}
where $conf$ = $\alpha$, $\beta$, $\gamma$, $\delta$, or $\epsilon$, 
$Exp_i$ is the number of counts in the $ith$ interval of the experimental 
excitation energy spectrum, and $C^{conf}_i$ is the number of counts 
in the same interval as predicted by the normalized peak configuration $conf$.
The $\chi^2$ tests were performed in the interval 
-1.515 MeV $<$ Excitation Energy $<$ 3.735 MeV, corresponding to 35 degrees
of freedom.   
%%The degree of freedom of the $\chi^2$ tests for all the configurations was 
%%conservatively chosen as 35, avoiding to perform the tests in the intervals 
%%where the number of counts was negligible.

Because of the properties of the subtraction of radiative 
effects from spectra quoted above,   
all the peak configurations $\alpha$, $\beta$, $\gamma$, 
$\delta$, and $\epsilon$  produce the same ``radiatively corrected''
spectrum. The spectra of 
Fig.~\ref{Monte_Carlo_Spectra Without_Radiative_Effects}
are obtained from 
Fig.~\ref{Monte_Carlo_Spectra With_Radiative_Effects}  by
turning off the radiative effects, that is replacing 
the  ``Fig.~\ref{Single_Peaks} red curve-like'' peaks with 
``Fig.~\ref{Single_Peaks} blue curve-like'' peaks, without changing positions 
and amplitudes of the peaks. All plots of 
Fig.~\ref{Monte_Carlo_Spectra Without_Radiative_Effects} are quite equal,
as confirmed by the $\chi^2$ test,
\begin{equation}
\chi^2 = \sum_{i}^{}{\frac{(C^{conf_1}_i-C^{conf_2}_i)^2}{C^{conf_1}_i}}
\; ,
\label{eq:17}
\end{equation}
with $conf_1$ and $conf_2$ = $\alpha$, $\beta$, $\gamma$, $\delta$, or 
$\epsilon$. In the worst case ($conf_1 = \alpha$ and $conf_2 = \gamma$) 
Eq.~(\ref{eq:17}) yielded a value of 28.387 with 40 degrees of freedom.  

In Fig.~\ref{Histogram_Comparisons}(a),
Fig.~\ref{Monte_Carlo_Spectra Without_Radiative_Effects}(a) (dashed line) 
and Fig.~\ref{Monte_Carlo_Spectra Without_Radiative_Effects}(c) 
(continuous line)
are shown together. Because the configurations $\alpha$ and 
$\gamma$ produce the two most different ``radiatively corrected'' SIMC 
results, the difference between the two curves plotted in 
Fig.~\ref{Histogram_Comparisons}(a) was chosen as the systematic error due 
to the ambiguity of the peak structure underlying the energy spectrum.  
As shown in Fig.~\ref{Histogram_Comparisons}(b)
this error is small compared to the statistical error. 

The method to obtain radiative corrected spectra described in this Appendix 
was used, in the analysis of the $^9_{\Lambda}$Li spectrum, because of the 
difficulties in establishing the peak structure underlying the experimental 
spectrum. It is relatively new and it could be worthwhile hence to make some 
considerations about its reliability. The method relies on Eq. (\ref{eq:14}) 
that is mathematically correct. The uncertainties on the radiative 
corrected spectrum $\overrightarrow{S}$ derived by Eq. (\ref{eq:14}) originate 
obviously from the uncertainties on the experimental spectrum 
$\overrightarrow{Exp}$ and on the function $R^{-1}$ ("detector function" in 
the following) that provides $\overrightarrow{S}$ once $\overrightarrow{Exp}$
has been measured. If $\overrightarrow{Exp}$ and $R^{-1}$ were exempt from 
errors $\overrightarrow{S}$ would be "perfect". To understand the effects on
the reconstructed radiative spectrum of the uncertainties on the measured 
spectrum and on the detector function it could be worthwhile to look at the 
results of the method to derive a neutron energy spectrum from the proton
recoil energy measurement (see for example \cite{Baba, Donzella}). 
This method consists in determining a neutron energy spectrum measuring 
the energies of the protons generated by scatterings of the neutrons 
in a radiator and is formally similar to the one described in 
this Appendix to derive radiative corrected spectra from the experimental ones. 
Formally, the connection between the neutron energy spectrum and the proton 
recoil energy spectrum can be expressed by a formula like Eq. (1) quoted in 
Ref. \cite{Baba} that can be concisely expressed as:

\begin{equation}
F(E_n)dE_n = D^{-1} \cdot Y(E_p)dE_p.
\label{eq:RecProt1}
\end{equation}

Here $F(E_n)dE_n$ is the number of neutrons in the neutron spectrum with an 
energy  included in the interval $ E_n -dE_n < E_n < E_n +  dE_n$,
$Y(E_p)dE_p$ is the number of  protons in the experimental spectrum with an 
energy $E_p -dE_p < E_p < E_p +  dE_p$, and $D^{-1}$ is 
the "detector function". Defining $\overrightarrow{N}$ and 
$\overrightarrow{P}$ the arrays whose elements are $F(E_n)dE_n$ and 
$Y(E_p)dE_p$ respectively ($E_n$ and $E_P$ covering the whole
neutron and proton spectra), Eq. (\ref{eq:RecProt1}) transforms into: 

\begin{equation}
\overrightarrow{N}  = D^{-1} \cdot \overrightarrow{P}.
\label{eq:RecProt2}
\end{equation}

Eq. (\ref{eq:RecProt2}) is formally equivalent to Eq. (\ref{eq:14})
(in Eq. (\ref{eq:RecProt2}) $D^{-1}$ is a diagonal matrix). 

However, in Eq. (\ref{eq:RecProt2}) the knowledge on $\overrightarrow{P}$ is 
(sometimes greatly) affected by  the uncertainties on the measured recoil 
proton energy. These uncertainties mean that the determination of the number of 
protons $Y_pdE_p$ whose real energy is included in the $ith$ interval of 
the proton energy spectrum $\overrightarrow{P}$
and hence the  proton energy spectrum itself are affected by 
(sometimes not negligible) uncertainties too.
Several factors affect the proton energy measurement: detector calibration, 
background subtraction, and, above all, proton energy losses in the detector 
elements (including their entrance windows) and in the air between them. 
To correct for proton energy losses, the proton energies are shifted by the 
estimated average energy loss over possible proton paths, or, sometimes,
in low energy regions, by unfolding techniques. The proton energy losses set 
usually the low limit of the reconstructed neutron energy spectrum. 
The detector function $D^{-1}$ in Eq. (\ref{eq:RecProt2}) is affected by 
uncertainties too. It includes the detector efficiency that depends on the 
geometry (and on the connected problem of the determination of the scattering 
angle of the detected proton) and on the differential n - p scattering cross 
section which are  both sources of systematic errors. The differential n - p 
scattering cross section is often obtained by parameterizations. Despite 
these problems, the method to derive a neutron energy spectrum from the proton 
recoil energy measurement provides usually satisfactorily results. In 
Ref \cite{Baba}, the method was applied to determine the spectra of nearly 
monoenergetic neutrons form the reaction $^7$Li(p,n)$^7$Be measured for eight 
incident proton energies. The situation was here complicated by the fact that, 
together with the neutrons generated in the reaction under study, which 
corresponded to a well-defined peak in the neutron energy spectrum, the 
experiment detected neutrons by other 
reactions, as the three-body breakup process $^7$Li(p,n$^3$He)$\alpha$, that 
generated a long tail in the low energy region of the neutron energy spectrum. 
Despite that, the reproduction of the peaks of the eight neutron energy spectra
was excellent, while the neutron counts in the tails of these spectra was 
somehow bigger than the corresponding parts of the spectra obtained with a 
Time Of Flight (TOF) detector for three of the eight incident proton energies. 
The authors decided to rely on the TOF detector results for the tails of these 
three spectra because the TOF detector was free from the problems concerning 
the effects of the proton energy loss by reactions in the detectors and 
becuase it extended to lower energies than the method based on the proton 
recoil energy measurement. In Ref. \cite{Donzella} the situation was improved 
with respect to Ref.  \cite{Baba} because the detector function $D^{-1}$ was 
simply equal, for all the proton energies, to  $\frac{1}{\cos^2(\theta)}$, with
$\theta$, the proton scattering angle, measured by two  silicon strip detectors
for the most energetic protons and, less precisely, through
the coordinates of the conversion point of the neutron inside the converter 
and the coordinates of the silicon detector closer to the converter for 
the protons whose energies were not big enough to make them reach the other 
silicon strip detector. The use of a segmented converter decreased the 
uncertainties on the proton energies due to energy losses inside the converter 
itself. As a consequence, the minimum neutron energy detectable was lower than
the one of Ref. \cite{Baba}. The double differential neutron 
yield for the reaction $^{13}$C(d,n) at 40 MeV was obtained this way.    
The good successes obtained in determining neutron spectra from the 
measurements of the proton recoil energies and the understanding of the 
effects that could make this method less effective ensure that we can rely on 
the method described in this Appendix to obtain, in the experiments performed 
at JLab, radiative corrected spectra from experimental binding energy spectra 
of hypernuclei. In fact, in these experiments, the hypernuclei are generated in
point-like targets and the experimental binding energy spectra are obtained by 
the measurements of the momenta and the scattering coordinates of the secondary
electrons and of the produced kaons in the reaction $Z(e,e^{\prime}K^+)Z-1$. 
These momenta and scattering coordinates are determined very precisely by 
magnetic spectrometers (at the level of $10^{-4}$ for the momenta measured by 
HRS arms). The energy losses of the secondary electrons and 
of the produced kaons inside the target and along the path 
to the detectors are small. Moreover, thanks to 
the excellent Particle Identification apparatus employed, the experimental 
spectrum of the experiment E94-107 was pratically background free, the only 
small background coming from kaons from accidental coincidences. The 
uncertanties on $\overrightarrow{Exp}$ of Eq. (\ref{eq:14}) are hence much 
smaller than the uncertanties on $\overrightarrow{P}$ 
of Eq. (\ref{eq:RecProt2}). Besides, the uncertainties on the detector 
function $R^{-1}$ in Eq. (\ref{eq:14}) are smaller than the 
uncertainties on the corresponding function $D^{-1}$ in 
Eq. (\ref{eq:RecProt2}) because of the simpler geometries 
involved, the smaller uncertainties on the scattering angles of the detected
particles and of the QED cross sections involved in the function $R^{-1}$  
better known than the neutron - proton cross sections involved in the function 
$D^{-1}$.

\section{Missing-mass scale}
\label{Appendix B}

In the Hall A experimental setup, scattered electrons and produced kaons of 
the reactions $^{9}$Be$(e,e'K^+)$\lamb{9}{Li} and 
$^{12}$C$(e,e'K^+)$\lam{12}{B}
were detected by the High Resolution Spectrometer (HRS) electron arm and 
by the HRS hadron arm, respectively, while the 
primary electrons were provided by the CEBAF accelerator. The CEBAF 
accelerator electron beam energy and the central momenta and angles of 
the HRS electron and hadron arms were set according to the kinematics 
of the reactions and are taken as constant for the course of the 
experiment (their variations being of the order of 
$10^{-5}$ for the CEBAF electron beam energy and the central momenta 
of the HRS electron and hadron arms, and practically zero for the spectrometer 
central angles). However, the actual values of the CEBAF accelerator
electron beam energy and of the central momenta and angles of the HRS 
electron and hadron arms, although constant, differ by unknown amounts 
from the nominal set values, and are referred to as ``kinematical 
uncertainties''.  
Although small (the experimental uncertainties on  the CEBAF accelerator
electron beam energy and on the spectrometer central momenta
being of the order  of $10^{-4}$ - $10^{-3}$ and 
those on the spectrometer central angles of the order of $10^{-2}$),  
these kinematical uncertainties cause a global shift in the binding-energy 
spectrum that hence has to be calibrated.  
In fact, the binding energy is expressed as
\begin{equation}
E_{bind} = -\sqrt{(E_m)^2-(P_m)^2} + M_{residue} + M_{\Lambda}\; ,
\label{eq:18}
\end{equation} 
with
\begin{equation}
E_m=M_{Target}+E_e-E_{e'}-E_K\; ,
\label{eq:19}
\end{equation}
and
\begin{equation}
\vec{P}_m=\vec{P}_e-\vec{P}_{e'}-\vec{P}_K\; ,
\label{eq:20}
\end{equation}
where $E_e$, $\vec{P}_e$, $E_{e'}$, $\vec{P}_{e'}$, $E_K$, and $\vec{P}_K$
are the energies and the momenta of the primary electron, of the scattered
electron and of the produced kaon respectively, $M_{\Lambda}$ is the 
$\Lambda$ mass, $M_{Target}$ is the target mass, and
$M_{residue}$ is the mass of the residual nucleus, that is of the nucleus with
A-1 nucleons and Z-1 protons (where A is the number of nucleons and Z is the 
atomic number of the target, respectively).

The change in the binding-energy 
spectrum $\Delta E_{bind}$ caused by the kinematical uncertainties 
can be expressed by the formula
\begin{eqnarray}
\Delta E_{bind}& = & \frac{\partial E_{bind}}{\partial E_e}\cdot \Delta E_e   
+  \frac{\partial E_{bind}}{\partial P_{E\_Arm}}\cdot \Delta P_{E\_Arm} 
\nonumber \\
& + & \frac{\partial E_{bind}}{\partial P_{H\_Arm}}\cdot \Delta P_{H\_Arm}
+\frac{\partial E_{bind}}{\partial \phi_{E\_Arm}}\cdot \Delta \phi_{E\_Arm}
\nonumber \\
& + & \frac{\partial E_{bind}}{\partial \phi_{H\_Arm}}\cdot \Delta 
\phi_{H\_Arm}\; ,
\label{eq:21}
\end{eqnarray}
where $E_e$ is the CEBAF electron beam energy, $P_{E\_Arm}$ and 
$P_{H\_Arm}$ are the central momenta of HRS electron 
arm and HRS hadron arm, $\phi_{E\_Arm}$ and $\phi_{H\_Arm}$ are the 
central angles (defined as the angles between the central axes and the 
CEBAF beam line) of HRS electron arm and HRS hadron arm, and
$\Delta E_e$, $\Delta P_{E\_Arm}$, $\Delta P_{H\_Arm}$, 
$\Delta \phi_{E\_Arm}$ and $\Delta \phi_{H\_Arm}$ are 
the kinematical uncertainties, that is the differences 
between the nominal and the actual values of the CEBAF electron beam 
energy and of the central momenta and the central angles of HRS electron 
and HRS hadron arms. $\vec{P}_e$,  $E_{e'}$, $\vec{P}_{e'}$, $E_K$, and 
$\vec{P}_K$ in Eq.~(\ref{eq:18})
-  (\ref{eq:20}) are functions of $E_e$, $P_{E\_Arm}$, $P_{H\_Arm}$, 
$\phi_{E\_Arm}$ and $\phi_{H\_Arm}$, as well as 
(with the exception of  $\vec{P}_e$) of the scattering variables. 
For example, the components normal to the beam line in the horizontal plane
of $\vec{P}_{e'}$ and $\vec{P}_K$ respectively
(called here $P_{e'_x}$ and $P_{K_x}$, respectively) are given by the expressions
$P_{e'_x}=P_{E\_Arm}\cdot(1+\delta_{e'})\cdot \cos(\theta_{e'})\cdot 
\sin(\phi_{E\_Arm} + \phi_{e'})$ and $P_{K_x}=P_{H\_Arm}\cdot(1+\delta_{K})
\cdot \cos(\theta_{K})\cdot \sin(\phi_{H\_Arm} + \phi_{K})$, 
while $E_{e'}=P_{E\_Arm}\cdot(1+\delta_{e'})$ and
$E_{K}=\sqrt{(P_{H\_Arm}\cdot(1+\delta_{K}))^2+m_k^2}$,
where  $\delta_{e'}$ ($\delta_K$) is the percentage 
difference between the momentum of the scattered electron (produced kaon) 
and the central momentum of the HRS electron arm (HRS hadron arm), 
$\phi_{e'}$ ($\phi_K$) is the angle between the 
electron (produced kaon) direction in the horizontal plane, and
 $\phi_{E\_Arm}$ ($\phi_{H\_Arm}$), $\theta_{e'}$ ($\theta_K$) 
are the angles between the electron (produced kaon) direction in the 
vertical plane and the CEBAF beam line, and $m_K$ is the kaon mass.

Considering Eqs.~(\ref{eq:18}), (\ref{eq:19}), and (\ref{eq:20}), 
Eq.~(\ref{eq:21}) is equal to
\begin{eqnarray}
 \Delta E_{bind} =  (-\Delta E_e+\Delta P_{E\_Arm}+\Delta P_{H\_Arm}) & &
\nonumber \\
 \times  \frac{M_{Target}+E_e-E_{e'}-E_K}
{\sqrt{(M_{Target}+E_e-E_{e'}-E_K)^2 - (\vec{P}_e-\vec{P}_{e'}-\vec{P}_K)^2}}
\nonumber \\
 + \frac{f}
{2\cdot
\sqrt{(M_{Target}+E_e-E_{e'}-E_K)^2 - (\vec{P}_e-\vec{P}_{e'}-\vec{P}_K)^2}} 
\nonumber \\
\label{eq:22}
\end{eqnarray}
where 
\begin{eqnarray}
f & = & \frac{\partial (\vec{P}_m)^2}{\partial E_e}\cdot \Delta E_e
+\frac{\partial (\vec{P}_m)^2}{\partial P_{E\_Arm}}\cdot \Delta P_{E\_Arm}
\nonumber \\
& + & \frac{\partial (\vec{P}_m)^2}{\partial P_{H\_Arm}}\cdot \Delta P_{H\_Arm}
+\frac{\partial (\vec{P}_m)^2}{\partial \phi_{E\_Arm}}\cdot \Delta 
\phi_{E\_Arm} \nonumber \\
& + & \frac{\partial (\vec{P}_m)^2}{\partial \phi_{H\_Arm}}\cdot \Delta 
\phi_{H\_Arm}
\label{eq:23}
\end{eqnarray}

 For the kinematics of the experiment E94-107 (see the nominal values of  
$E_e$, $E_{e'}$, and $E_K$ at the end of this Appendix) 
or if $M_{\textrm{Target}} \gg E_e, E_{e'}, E_K$,
the change in the binding-energy 
spectrum $\Delta E_{bind}$ caused by the kinematical uncertainties is 
\begin{equation}
\Delta E_{bind}\sim S + T\, ,
\label{eq:24}
\end{equation}
where
\begin{equation}
 S = -\Delta E_e + \Delta P_{E\_Arm} + \Delta P_{H\_Arm}\, ,
\label{eq:25}
\end{equation}
and
\begin{equation}
 T = M_{\textrm{Target}}^{-1}\cdot \frac{f}{2}\ .
\label{eq:26}
\end{equation}
The term $S$ does not depend on the target mass and causes a global 
shift of the binding-energy spectrum without changing the peak
shapes and relative positions, while the term $T$
is equal to zero only when $0\!=\!\Delta E_e\!=\!\Delta P_{E\_Arm}\!=\!
\Delta P_{H\_Arm}\!=\!\Delta \phi_{E\_Arm}\!=\!\Delta \phi_{H\_Arm}$. 
When this condition is not fulfilled, the main effect of $T$
on the binding-energy spectrum is to broaden the peaks
because, in this case, depending on scattering variables through $\vec{P}_m$,
it produces non-zero, scattering variable dependent values of  
$\Delta E_{bind}$ and, as a consequence, an unphysical dependence on the 
scattering variables of the binding energy that broadens the peaks 
corresponding to the bound states.
Because of the presence of the coefficient $M_{\textrm{Target}}^{-1}$,
the effect of $T$ on the position of the peaks is negligibly small.

The actual values of  $E_e$, $P_{E\_Arm}$, $P_{H\_Arm}$, $\phi_{E\_Arm}$, 
and $\phi_{H\_Arm}$ are hence those for which $S\!=\!T\!=\!0$
(which places a peak at its known position in the binding energy
spectrum and minimizes its FWHM). 

To determine the $\Lambda$ binding energy of the hypernucleus 
\lamb{9}{Li} produced in the reaction  $^{9}$Be$(e,e'K^+)$\lamb{9}{Li},
we took advantage of the fact that the experiment used the same 
set-up employed for the study of the \lam{12}{B} hypernucleus.
The values of  $E_e$, $P_{E\_Arm}$, $P_{H\_Arm}$, $\phi_{E\_Arm}$, and 
$\phi_{H\_Arm}$ were thus the same in the productions of both
\lamb{9}{Li} and \lam{12}{B} and were determined by positioning 
the \lam{12}{B} ground state at its known position of $11.37\pm 0.06$ 
MeV~\cite{Carbon_Ground_State} in the binding-energy spectrum and 
minimizing its FWHM. 
When minimizing the  \lam{12}{B} ground-state FWHM, it has to be taken 
into account that this ground state is actually a doublet whose
energy splitting, assumed to be equal to the \lam{12}{C} ground-state
energy splitting, is 161.6 $\pm$ 0.2 keV~\cite{Ma, Hosomi}. 
Minimizing the \lam{12}{B}
ground-state FWHM implies hence some sort of distortion because it 
artificially narrows the positions of the peaks making up the doublet.
However, the \lam{12}{B} ground-state doublet energy 
splitting is small enough with respect to the energy resolution of the
experiment to make the approximation of assuming the \lam{12}{B} ground state
as a single peak still valid. No attempt to
minimize the FWHM was performed on the other peaks of the \lam{12}{B}
binding-energy spectrum. 
Another possibile source of distortion comes from the term $T$ in 
Eq. (\ref{eq:24}), which, although small because of the presence of the factor 
$M_{Target}^{-1}$ in it (see Eq. (\ref{eq:26})), 
can potentially change the positions
of the excited states with respect to each other and with respect 
to the ground state. During all the process of minimization of the 
kinematical uncertainties, the positions of the peaks of the 
\lam{12}{B} energy spectrum as resulted by a fitting procedure were checked
to ensure that the relative peak positions did not change within the errors
(the error of a position peak being defined as the standard deviation
resulting by the fitting procedure). 
 It has to be stressed that the term $S$ in the 
expression of $\Delta E_{bind}$ given by Eq.~(\ref{eq:24})
dominates because of the presence of the factor $M_{Target}^{-1}$ in the 
term $T$ (see Eq.~(\ref{eq:26})). The positioning of the \lam{12}{B} 
ground state at its known value in the binding spectrum was thus mainly 
performed choosing a set of values $\Delta E_e$, $\Delta P_{E\_Arm}$, 
and $\Delta P_{H\_Arm}$ that produced
a value of $S$ equal to the difference between the measured and the 
expected position of the \lam{12}{B} ground state. The minimization of 
the  \lam{12}{B} ground state peak FWHM produces only second order effects 
on the position of the peaks in the binding-energy spectrum and was
peformed mainly to choose the right set of values
$\Delta E_e$, $\Delta P_{E\_Arm}$, and  $\Delta P_{H\_Arm}$ among the
$\infty^2$ sets of values that produced the desired value of $S$. 
Things are much different for the reaction $p(e,e'K^+)\Lambda$,
where the target mass is small and the minimization of the FWHM of the peak 
(which is a single peak) plays a role as important as the positioning 
of the peak in the binding-energy spectrum to its zero value.

The procedure described above resulted in the  set of values 
$E_e\!=\!3775.38$ MeV, $P_{E\_Arm}\!=\!1573.63$ MeV, 
$P_{H\_Arm}\!=\!1955.79$ MeV, $\phi_{E\_Arm}\!=\!-5.940^{\circ}$, and 
$\phi_{H\_Arm}\!=\!6.050^{\circ}$. Replacing with these values the 
nominal ones $E_e\!=\!3774.96$ MeV, $P_{E\_Arm}\!=\!1570$ MeV, 
$P_{H\_Arm}\!=\!1960$ MeV, $\phi_{E\_Arm}\!=\!-5.873^{\circ}$, and
$\phi_{H\_Arm}\!=\!6.131^{\circ}$) in the $E_{bind}$ expression, a value 
of $8.36 \pm 0.08$ (stat.) MeV was obtained for the \lamb{9}{Li} ground state. 

The statistical error of $\pm 0.08$ MeV is the error in the position of 
the first peak in the four-Gaussian fit of the \lamb{9}{Li} binding-energy
spectrum (see Fig.~\ref{databe} and Table~\ref{tab:results}). When evaluating
the systematic error of the \lamb{9}{Li} binding energy, one has to 
consider that, as quoted above, the term $S$ in $\Delta E_{bind}$ dominates  
the effects of the term $T$. When neglecting the energy loss in the target, 
this means that if the single 
values of $\Delta E_e$, $\Delta P_{E\_Arm}$, and $\Delta P_{H\_Arm}$  
were wrong, the  \lamb{9}{Li} binding energy would not be affected 
significantly as long as the sum $S$ (and hence the position of the 
\lam{12}{B} ground state) is correctly reproduced. The major source of 
systematic error is hence due to the energy loss in the target. The 
difference between the shifts of the ground-state position in the 
\lam{12}{B} and  \lamb{9}{Li} binding-energy spectra due to the 
energy loss in the targets of $^{12}$C and $^9$Be, respectively, was evaluated 
to be equal to $50$ keV through the use of the Monte Carlo code 
SIMC~\cite{SIMC}.
This value was added in quadrature to the error of $60$ keV quoted 
for the \lam{12}{B} ground-state binding energy~\cite{Carbon_Ground_State}
to give a $80$ keV systematic error on the \lamb{9}{Li} binding energy. 

\begin{acknowledgments}
We want to remember F. Cusanno, who enthusiastically and greatly contributed to 
this paper before his untimely death. 
We acknowledge the Jefferson Lab Physics and Accelerator
Division staff for the outstanding efforts that made
this work possible. 
%This work was supported by U.S. DOE contract
%DE-AC05-84ER40150, Mod. nr. 175, under which the Southeastern
%Universities Research Association (SURA) operates the Thomas
%Jefferson National Accelerator Facility, 
This material is based upon work supported by 
the Department of Energy, Office of Energy Research, 
under contract DE-AC05-06OR23177,
by the Italian Istituto
Nazionale di Fisica Nucleare, by the Grant Agency of the
Czech Republic under grant No. P203/12/2126, by the French CEA
and CNRS/IN2P3, and by the U.S. National Science Foundation.
\end{acknowledgments}

\end{document}